\documentclass[ ]{aa}
\usepackage{color}
\usepackage{natbib}
\bibpunct{(}{)}{;}{a}{}{,} 

\def\msun{$\mbox{M}_\odot$}

\def\afe{{[$\alpha$/Fe]}}
\def\feh{{[Fe/H]}}

\begin{document}

\title{Self-similarity in the chemical evolution of galaxies and the delay time distribution of SNe Ia}
\subtitle{}

\author{C.J.~Walcher\inst{\ref{i:CJW}}; R.M.~Yates\inst{\ref{i:rob1},\ref{i:rob2}}; I.~Minchev\inst{\ref{i:CJW}}; C.~Chiappini\inst{\ref{i:CJW}}; M.~Bergemann\inst{\ref{i:marie}};  G.~Bruzual\inst{\ref{i:GB}}; S.~Charlot\inst{\ref{i:SC}}; 
P.R.T.~Coelho\inst{\ref{i:PC}}; A.~Gallazzi\inst{\ref{i:AG}}; M.~Martig\inst{\ref{i:marie}}}

\institute{
Leibniz-Institut f\"ur Astrophysik Potsdam (AIP), An der Sternwarte 16, 14482 Potsdam, Germany \label{i:CJW}
\and
Max Planck Institut f\"ur Extraterrestrische Physik, Giessenbachstra{\ss}e 1, 85748 Garching, Germany \label{i:rob1}
\and
Max Planck Institut f\"ur Astrophysik, Karl-Schwarzschild-Str. 1, 85748 Garching, Germany  \label{i:rob2}
\and
Max Plank Institut f\"ur Astronomie, K\"onigstuhl 17, 69917 Heidelberg, Germany  \label{i:marie}
\and
Instituto de Radiastronomia and Astrofisica (IRyA), Morelia, Michoacan 58089, Mexico \label{i:GB}
\and
Sorbonne Universit\'es, UPMC-CNRS, UMR7095, Institut d'Astrophysique de Paris, F-75014, Paris, France \label{i:SC}
\and
Instituto de Astronomia, Geofisica e Ciencias Atmosf\'ericas, Universidade de S\~ao Paulo, Rua do Matao 1226, 05508-090 - S\~ao Paulo - Brasil \label{i:PC}
\and
INAF-Osservatorio Astrofisico di Arcetri, Largo Enrico Fermi 5, I-50125 Firenze, Italy \label{i:AG}
}

\authorrunning{Walcher, et al.}
\titlerunning{Selfsimilar cheevo}
\date{Received date / Accepted date }

\abstract{Recent improvements in the age dating of stellar populations and single stars allow us to study the ages and abundance of 
stars and galaxies with unprecedented accuracy. We here compare the relation between age and $\alpha$-element abundances for 
stars in the solar neighborhood to that of local, early-type galaxies. We find both relations to be very similar. Both fall into 
two regimes with a flat slope for ages younger than $\sim$ 9 Gyr and a steeper slope for ages older than that value. This quantitative 
similarity seems surprising, given the different types of galaxies and scales involved. For the sample of early-type galaxies we also 
show that the data are inconsistent with literature delay time distributions of either single or double Gaussian shape. The data are 
consistent with a power law delay time distribution. We thus confirm that the delay time distribution inferred for the Milky Way from chemical 
evolution arguments also must apply to massive early-type galaxies. We also offer a tentative explanation for the seeming universality 
of the age-\afe~relation as the manifestation of averaging of different stellar populations with varying chemical evolution histories. }


\maketitle

\section{Introduction}
\label{s:intro}

It has long been recognized \citep{tinsley79,matteucci86} that the element abundance ratio \afe~is a powerful estimator of the duration 
of star formation events in galaxies. This is because of the different explosion timescales and yields of different types of supernovae. 
A direct consequence of this insight is the expectation of a correlation between the ages of stars in galaxies and their \afe~ratios. 
A recent example is Figure 1 in \citet{chiappini15}, which shows the generic prediction for single stars in the Milky Way. 
It is unclear, however, how this relation would translate into galaxy-wide average properties. Generally, 
one expects that galaxies that have stopped forming stars at an earlier time in the history of the universe (equivalent 
to having a shorter star formation timescale), would show a smaller contribution of light from Fe-enriched stars in their spectra and 
would thus show a higher overall \afe~enrichment.  There does not seem to be a good reason why the relations between age 
and \afe~ should be quantitatively the same for entire galaxies and single stars, with the hope that possible differences could be used to 
study the different star formation histories. However, the exploration of this 
expected correlation has been hampered by uncertainties in stellar and galaxy ages, related to both model uncertainties and to intrinsic 
degeneracies, such as the age-metallicity degeneracy.

We have recently been able to take a significant step forward by showing the existence of this correlation for early-type galaxies (ETGs)
\citep[][hereafter W15]{walcher15}. Indeed, earlier work such as \citet{jorgensen99} found no correlation between \afe~and age. 
The first time this correlation was tentatively seen is by \citet{gallazzi06}. A correlation of age and \afe~was unambiguously shown by 
\cite{graves10} from stacked spectra (their Fig.~4), but the very nature of stacked spectra made it impossible to study the scatter in the 
relation. The relation was shown on a per galaxy basis by \cite{kuntschner10} (their Fig.~6), but in this case small sample size and continued  
large uncertainties on age made an interpretation difficult. Other recent work, such as \citet{thomas10} and \citet{johansson12} show and 
discuss the parameters age, \feh~and \afe, but do not directly address the age-\afe~relation explored here. The W15 results are 
nevertheless qualitatively in agreement with these earlier 
papers and reinforce and expand on them. We emphasize that for this same correlation, it is important to heed the warnings of 
\cite{thomas05}, who discuss the importance of degeneracies when using age as a parameter. We quantitatively show in 
W15 that the age-metallicity degeneracy does not give rise to the observed correlation. 

An interesting parallel development has been the verification of the expected similar correlation in the stars of the Milky Way. 
The unique age-metallicity relation in the Galactic disk has been first suggested by \citet{twarog80} using multi-band photometry 
data. However, \citet{edvardsson93} and later studies \citep{feltzing01, nordstrom04, holmberg07, holmberg09, 
casagrande11} have found that there is no one-to-one relationship between ages and metallicities of stars and a large scatter 
at any age that may have an astrophysical cause. Finally, the most recent work by \citet[][hereafter B14]{bergemann14}, using the 
high-resolution spectra from the Gaia-ESO stellar survey, has conclusively established the weak age-metallicity relation in the solar 
vicinity of the Galactic disk. This is the first study to carefully analyze the survey target selection effects and their impact on the age - metallicity 
diagram. For the stars with ages below 8 Gyr and for the solar vicinity, the observed age-metallicity relation was found to be nearly 
flat, and the majority of older stars turned out to be metal-poor and enhanced in $\alpha$ elements. Similar conclusions were reached 
by \citet[][hereafter H13]{haywood13} and \citet{bensby14}. As discussed in \citet{bensby14}, the H13 analysis lead to a very 
tight \afe-age relation due to the problems of the spectroscopic analysis and sample selection biases. Generally, B14 
established that \afe~is a good proxy for the age of a star, even though they see a significant dispersion of [Mg/Fe], especially 
at ages above 9 Gyr. 

This paper attempts to establish two new statements. First, the correlation between age and \afe~as expected from chemical 
evolution is seen in ETGs and is quantitatively similar to the one for stars in the solar neighborhood. This is true despite 
the very different star formation histories of these two different kind of stellar systems. Second, this universality allows to explore the 
dependence on the yields and delay time distributions of SNe Ia and II. When fixing the yields, the 
age-\afe~relation of ETGs thus provides additional interesting constraints on the delay time distribution of SNe Ia. 

\section{Data and models }
\label{s:sources}

We are interested in comparing the relation between age and \afe~for galaxies and stars and for data and models. We here describe the 
data and models we use for the present contribution. 

\subsection{Data for ETGs}
\label{s:W15}

For observational data concerning galaxies we turn to our publication of W15. There we analyzed a spectroscopic sample of 
2286 ETGs selected from the SDSS survey, data release 7 \citep{abazajian09}. The galaxies were selected to show no 
emission lines (and therefore no visible star formation), to be photometrically concentrated, and to have yielded spectra with 
sufficient signal-to-noise (S/N$>40$) to allow a careful analysis of the stellar population content. To analyze the spectra we used 
the pixel fitting code {\tt paradise}. This algorithm fits a linear combination of simple stellar populations to the galaxy data, at the same 
time as deriving the optimal kinematic parameters velocity and velocity dispersion. The stellar population models used were the 
differential stellar population models of \citet{walcher09}. 
In particular we derived the physical parameters age, \feh, and \afe. In the present contribution we only use those physical properties 
as derived in a luminosity-weighted sense, i.e. every stellar population contributes to the total signal according to its luminosity contribution to 
the overall spectrum. In W15 we also addressed the ability to actually separate the properties of the old and intermediate age stars on a per 
galaxy basis. Typical errorbars (precision) on age are 0.2 Gyr and 0.01 dex on \feh~and \afe\footnote{We remind the reader that these age 
precisions are obtained for stellar populations, i.e. averages of many stars. The techniques used to derive these ages are very different from 
the techniques used for single stars.}. The definition of the $\alpha$-element abundances 
groups together the elements O, Ne, Mg, Si, S, Ca and Ti \citep{coelho07}, but the dominant signal in the wavelength range we use for 
determination of the abundance will come from Mg. The models are normalized to the solar abundances from \citet{grevesse98}. 
The galaxies cover a mass range from $10^{10.2}$ to $10^{11.5}$ \msun. 

\subsection{Data for Milky Way stars}
\label{s:H13}

For data on stars we turn to the publications of B14 and H13. First, we use the data from the Gaia-ESO spectroscopic 
survey, presented in B14. The Gaia-ESO survey \citep{gilmore12, randich13} is a large high-resolution spectroscopic 
survey of FGK stars in the Milky Way disk to date. The B14 dataset consists of 144 stars with ages from 0.5 to 13.5 Gyr, which were 
determined consistently using state-of-the-art stellar evolution models \citep{serenelli13}, and carefully verified on the accurate 
seismic estimates for the reference benchmark stars \citep{jofre14, heiter15}. The chemical abundances of 15 elements were determined 
using the high-resolution (R $\sim$ 47\,000) Gaia-ESO UVES spectra using the MARCS model atmospheres and experimental 
atomic line lists. The mean uncertainties are 1.5 Gyr in age, and 0.06  dex in metallicity and chemical abundances of 
$\alpha$-elements. The stars in the sample are all within 6 kpc to 9.5 kpc from the Galactic centre and are located close to 
the plane, |Zl $<$ 1.5 kpc.

Second, we use the data from the publication of H13. These authors published ages for single stars with known \feh, and \afe~
in the solar neighborhood. Their sample is based on the HARPS GTO observations of 1111 stars as published in \citet{adibekyan12}. 
The original sample had to be severely pruned to 363 stars with robust ages. This down-selection was based on an absolute magnitude 
cut at M$_{V} < 4.75$ and on a somewhat less reproducible selection of stars with "a well defined probability function" (H13). 
H13 note that their absolute age scale could be off by 1 to 1.5 Gyr, while relative ages would have uncertainties of 1 Gyr. 
The H13 definition of \afe~includes the mean of Mg, Si, and Ti abundances. In the analysis of W15 the Mg$_{\mathrm{b}}$ feature will dominate, 
therefore these two observational definitions are very comparable despite the different definition of the $\alpha$ group. 

The stellar data for H13 were read off Figure 6 and 17 using the {\tt PlotDigitizer} application. We were able to read off 112 
points in Figure 6 (age vs. \afe) and 300 points in Figure 9 (age vs.~\feh). The larger number of points in the age vs.~\feh~
plane is caused by the larger scatter, making it possible to distinguish more data points in the figure. As we are not 
interested in the properties of single stars but in the slopes and zero points of the correlations, we expect little bias if any 
from this sample incompleteness. In particular for the age vs.~\afe~relation, most of the invisible (crowded) points seem 
to be concentrated at low ages and low \afe, right on the general trend. Including the whole sample would thus presumably 
mainly decrease the scatter around this mean relation, but \emph{not} change the parameters of the relation. 

Since the genesis of this paper, more samples have appeared that extended the very local samples used here using CoRoT 
and Kepler data with spectroscopic follow-up \citep{chiappini15, anders16, martig15}. 
Adding these stars would not change the conclusions of this paper in an way. 

\subsection{Semi-analytic models of ETG formation}
\label{s:Y13}

The galaxy models are based on the semi-analytic models described in \citet[][, hereafter Y13]{yates13}, which are themselves 
an update of the Munich semi-analytic model, L-GALAXIES \citep{springel01, guo11}. In a nutshell, the model is built on merger 
trees from the Millennium \citep{springel05} and Millennium-II \citep{boylan-kolchin09} N-body simulations of DM structure 
formation and uses an analytic treatment to track the transfer of mass between different baryonic components of a galaxy, 
such as bulge and disk stars, hot and cold gas, etc. Prescriptions for supernova and AGN feedback 
are included. The most important ingredients for the present contribution are those that directly influence the chemical evolution, 
i.e.~SN yields, initial mass function (IMF), stellar lifetimes etc. All of these are described in detail in Y13. 

The only parameter that we treat as a variable in the present contribution is the delay time distribution (DTD) of SNeIa. The DTD 
describes the probability for a SN Ia to explode as a function of the time elapsed since a star formation event. The overall explosion 
rate of SNe Ia in a galaxy will depend on the DTD and the star formation history. The Y13 paper considers three DTDs: power-law, 
Bi-modal and Gaussian. The bi-modal DTD could be reasonably close to a power-law DTD for a specific choice of parameters (normalisation, slope, characteristic time, etc.). Here we choose parameters that have been proposed in the literature based on observations of 
the SNIa rate, but that still keep the DTDs 
sufficiently unique that our data and model matching allow us to distinguish between them. Formal parameter minimization 
of different DTDs and further dependencies (such as a metallicity dependance of the DTD) will be explored in future work. 

The Y13 model provides the same parameters as for the W15 ETGs, i.e.~age, \feh~and \afe. 
Just as for the ETG data from W15, the ages are calculated as r-band luminosity weighted ages. The \afe~value used in Y13 is 
actually the value of [O/Fe] and it is normalized to the \citet{anders89} meteoric abundances (i.e.~[O/H]=8.93 and [Fe/H]=7.51). 
Normalizing to the \citet{grevesse98} abundances would shift the overall normalization down by 0.1 dex in \afe. 
We have also tested the effect of taking the average of the enhancements of O, Si, S, and Ca (i.e.~not including Mg) as our value 
for \afe. All results of this paper are independent of whether we use [O/Fe] or this restricted definition of \afe~for the model galaxies. 
We decided to avoid [Mg/Fe], because there are some known peculiarities with the yields of this element in the yield set used 
\citep{portinari98}. In particular uncertainties concern the greater Mg production in low metallicity stars as compared to high metallicity 
stars, due to complex assumptions about pre-SN stellar winds.

It is also important to note exactly which sample we are using. Indeed, to the basic set of model ellipticals from Section 6.3 of Y13 
we impose an overall lower-mass limit of log(M*)=10.0, in order to roughly match that of the W15 sample. Here, we did {\emph not} 
impose the additional cut based on the 1$\sigma$ scatter of the \citet{johansson12} mass-age relation (see Section 6.3.1 of Y13). 
This additional cut would have removed those low mass model galaxies that we know are too old and red, due to efficient stripping 
and SN feedback in the model causing these objects to have run out of star-forming gas very early. Low-mass galaxies are, however, 
not considered in the present contribution.

\subsection{Simulations of disk assembly}
\label{s:M13}

For a chemical evolution model of the solar neighborhood stars, closely matched to the B14 sample, we now turn to the work 
by \citet[][hereafter M13]{minchev13}\footnote{The Y13 model also predicts abundance trends for disk galaxies and could have been used in the 
same way. However the M13 model has been constructed specifically for the Milky Way and thus is most directly comparable to the B14 
data. Also, using two entirely different chemical evolution models reinforces our statement that the age-\afe~relation is universal. }. 
The M13 model in turn is based on a simulation in the cosmological context by \citet{martig12} and the interested reader is referred to 
that paper for all details on the method. The main point for our discussion being that M13 choose the one galaxy out of all \citet{martig12} 
galaxies that most resembles the Milky Way. The chemical evolution model is tied to the dynamic evolution by having both disks 
grow inside out, similar gas-to-stellar mass ratio, and resampling the star formation rate in the simulation to match that of the semi-analytical 
chemical model. This method allows the circumvention of problems with fully self-consistent chemo-dynamical simulations, 
which occur due to uncertainties in subgrid physics -- even in 
high-resolution cosmological simulations, one particle represents $10^4$-$10^5$ \msun. The M13 paper readily supplies 
the [Mg/Fe] abundances of the stars out of a total of $\sim$30 elements. The M13 model uses Mg as its proxy of the 
$\alpha$-element group, which is compatible with the W15 and B14 analyses. 

The H13 data are limited to the Hipparcos volume, while the B14 data cover a somewhat more extended solar vicinity. 
To reproduce the limited volume in the 
data we look at a ring at radius r=8 kpc, of radial width $\Delta$r=0.1 kpc, and of vertical 
height $\Delta$|z|=0.05 kpc. We convolve the model with ad-hoc, but realistic errorbars, namely $\Delta$ age=1 Gyr 
and $\Delta$ \afe=0.11. Out of the total sample of available stars in the model ($\sim10^5$) we selected 400 stars randomly, 
which is approximately the size of the B14 and H13 samples combined. 

There is an important feature to the model, which is that its oldest stars are 11.2 Gyr old (12.2 Gyr including fiducial errorbars). 
The oldest stars in the observations 
can be as "old" as 15 Gyr. Clearly there is a difference in age scale, which may be imputed both to the observations and the 
simulations, for different reasons. Observationally, age scales may be uncertain due to several reasons, as discussed in B14. 
In the simulations on the other hand, the major effect is that the model is a pure thin disk model, i.e.~a chemodynamical simulation 
that was run for 11.2 Gyrs.  In the two-infall model from \citep{chiappini97}, the thick disk does pre-enrich the thin disk. Nevertheless, 
thin disk stars are chemically nearly independent of the thick disk stars, the chemical clock is essentially reset at the beginning of the 
second infall. The use of M13 simulation is justified, because the thick disk population shown in Figure 1 of \citet{chiappini15} 
is not present in the observational samples used here, see Section \ref{s:H13}.  
The net effect is that in simulations chemical evolution starts at 11.2 Gyr instead of $\sim$13. 
For our application this has the effect that we need to \emph{stretch} the age axis for the simulations somewhat to match the 
chemical evolution patterns of the observed Milky Way. The stretch factor therefore should be of order $\sim$1.2. This stretch 
factor will be further discussed in Section \ref{s:obs_age_afe}. 

\section{Results}

\subsection{The age-\afe~relation}
\label{s:obs_age_afe}

In this section we plot and compare the relations between age and \afe~and \feh. As discussed above, the different datasets have 
to be set on the same scale before being directly comparable. We apply the following scaling factors: (1) None to the W15 data. 
(2) A downward shift to the \afe~value of Y13 of 0.1 dex, which is justified by the different solar abundances used as reference. 
No shift is applied to \feh, as the nominal shift of 0.01 dex is not significant for the present work. 
(3) None to the B14 / H13 data. (4) As justified in Section \ref{s:M13} a correction factor of order 1.2 is expected to be needed due to 
differences in timescale between the M13 model and the Milky Way data. In practice we find that the factor 1.17 works well for the 
self similarity arguments exposed here. We note that this ad-hoc stretch factor makes it impossible for the moment to use the M13 
models to infer information on the SNeIa delay time distribution. 

\begin{figure}[tbp]
\begin{center} 
\resizebox{1.0\hsize}{!}{\includegraphics[]{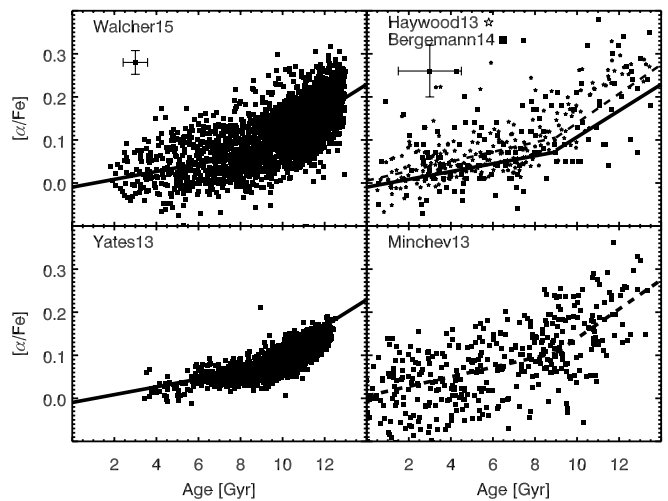}}
\end{center}
\caption[Comparison]{Comparison of the correlations between age and \afe~for two very different kinds of astrophysical objects. 
\emph{Upper left panel:} The luminosity weighted average properties of early-type galaxies from W15. The solid line is a formal 
fit to the two regimes, separated at 9 Gyr. \emph{Upper right panel:} Stars in the local neighborhood from B14 (solid squares) and 
H13 (stars). The dashed line is a formal fit to the B14 data for the two regimes, separated at 9 Gyr. The solid line repeats the fit for 
galaxies from the left panel. \emph{Lower left panel:} The luminosity weighted average properties of early-type galaxies 
in the semi-analytic model of Y13. The solid line repeats the fit for the W15 data for comparison. Note that no observational 
errors have been added to the model galaxy properties, which largely explains the difference in scatter. \emph{Lower right panel:} 
Single star properties for a simulated solar neighborhood from M13. The dashed line repeats the fit to the B14 data for comparison. 
Here, observational errors have been added for better comparison of scatter.}
\label{f:obs_alp}
\end{figure}

\begin{figure}[tbp]
\begin{center} 
\resizebox{1.0\hsize}{!}{\includegraphics[]{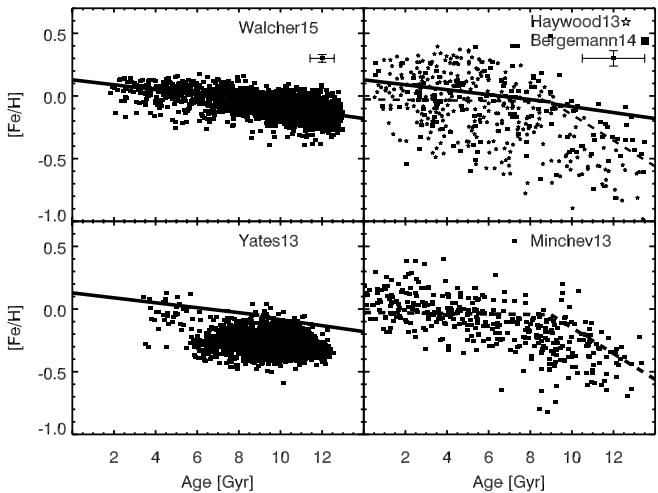}}
\end{center}
\caption[Comparison]{Comparison of the correlations between age and \feh~for two very different kinds of astrophysical objects. 
The panels and lines are the same as in Figure \ref{f:obs_alp}.}
\label{f:obs_feh}
\end{figure}

We plot the relation between age and \afe~in Figure \ref{f:obs_alp}. Qualitative agreement was expected from the literature on 
chemical evolution cited in Section \ref{s:intro}. Surprisingly, the relations are also quantitatively similar, all showing 
a clear change of slope at ages between 9 and 10 Gyr in both datasets and both model sets. On the other hand, the age-\feh~relations in 
Figure \ref{f:obs_feh}, while showing the overall same trend of \feh~decreasing with lookback time, are quantitatively very 
different in the sense that \feh~for old stars is much lower in the solar neighborhood. The galaxy data do not seem to require 
a two slope regime, whereas the stellar data do. We discuss possible reasons for this in Section 4.1.

We have verified that for all relations being studied here the Spearman-Rank test indicates that the probability of absence of any 
correlation is zero. We quantify the correlations by means of formal fits to each set of two parameters combinations using the 
{\tt LINFIT} module in IDL. We have fitted the two regimes separately and report the results in Table \ref{t:fitres}. For the 
age-\afe~relation all slopes are consistent at the 2$\sigma$ level (errorbars reported in the table are 1$\sigma$ errorbars). 
Likewise, all intercepts are the same within 2 $\sigma$, 
with the exception of the \feh~intercepts in the young regime. These last intercepts will depend more strongly on sample 
selection than any other, so we neglect this difference for the present contribution. 

\begin{table*}
\centering
\caption{ Coefficients of linear fits to the datasets  }
\begin{tabular}{ccc|cc}
Dataset & Parameters & Age range & Intercept  & Slope \\
\hline
W15 & age vs.~\afe & $<9$Gyr &  -0.010$\pm$0.0044 &  0.009$\pm$0.0006 \\
W15 & age vs.~\afe & $\ge9$Gyr & -0.199$\pm$0.0071 &  0.031$\pm$0.0006 \\
Y13 & age vs.~\afe & $<9$Gyr &  0.005$\pm$0.0045 &  0.008$\pm$0.0006 \\
Y13 & age vs.~\afe & $\ge9$Gyr & -0.197$\pm$0.0041 &  0.028$\pm$0.0004 \\
B14 & age vs.~\afe & $<9$Gyr &  0.005$\pm$0.0206 &  0.011$\pm$0.0037 \\
B14 & age vs.~\afe & $\ge9$Gyr & -0.200$\pm$0.0096 &  0.034$\pm$0.0011 \\
M13 & age vs.~\afe & $<9$Gyr & -0.007$\pm$0.0059 &  0.013$\pm$0.0012 \\
M13 & age vs.~\afe & $\ge9$Gyr & -0.151$\pm$0.0688 &  0.030$\pm$0.0065 \\ 
\hline
W15 & age vs.~\feh & $<9$Gyr &  0.129$\pm$0.0043 & -0.020$\pm$0.0006 \\
W15 & age vs.~\feh & $\ge9$Gyr &  0.151$\pm$0.0076 & -0.024$\pm$0.0007 \\
Y13 & age vs.~\feh & $<9$Gyr & -0.030$\pm$0.0247 & -0.028$\pm$0.0031 \\
Y13 & age vs.~\feh & $\ge9$Gyr & -0.065$\pm$0.0155 & -0.020$\pm$0.0015 \\
B14 & age vs.~\feh & $<9$Gyr & -0.027$\pm$0.0578 & -0.002$\pm$0.0103 \\
B14 & age vs.~\feh & $\ge9$Gyr &  0.882$\pm$0.0164 & -0.103$\pm$0.0018 \\
M13 & age vs.~\feh & $<9$Gyr &  0.104$\pm$0.0124 & -0.035$\pm$0.0025 \\ 
M13 & age vs.~\feh & $\ge9$Gyr &  0.308$\pm$0.1802 & -0.061$\pm$0.0170 \\
\hline
W15 & age vs.~\afe & $<9$Gyr & -0.010$\pm$0.0044 &  0.009$\pm$0.0006 \\
W15 & age vs.~\afe & $\ge9$Gyr & -0.199$\pm$0.0071 &  0.031$\pm$0.0006 \\
Y13 PL & age vs.~\afe & $<9$Gyr &  0.005$\pm$0.0045 &  0.008$\pm$0.0006 \\
Y13 PL & age vs.~\afe & $\ge9$Gyr & -0.197$\pm$0.0041 &  0.028$\pm$0.0004 \\
Y13 NG & age vs.~\afe & $<9$Gyr & -0.051$\pm$0.0080 &  0.010$\pm$0.0010 \\
Y13 NG & age vs.~\afe & $\ge9$Gyr & -0.629$\pm$0.0110 &  0.071$\pm$0.0010 \\
Y13 BM & age vs.~\afe & $<9$Gyr & -0.023$\pm$0.0039 &  0.004$\pm$0.0005 \\
Y13 BM & age vs.~\afe & $\ge9$Gyr & -0.137$\pm$0.0029 &  0.017$\pm$0.0003 \\
\end{tabular}
\label{t:fitres}
\end{table*}

\subsection{The Delay Time Distribution of SNeIa}
\label{s:obs_dtd}

\begin{figure}[tbp]
\begin{center} 
\resizebox{1.0\hsize}{!}{\includegraphics[]{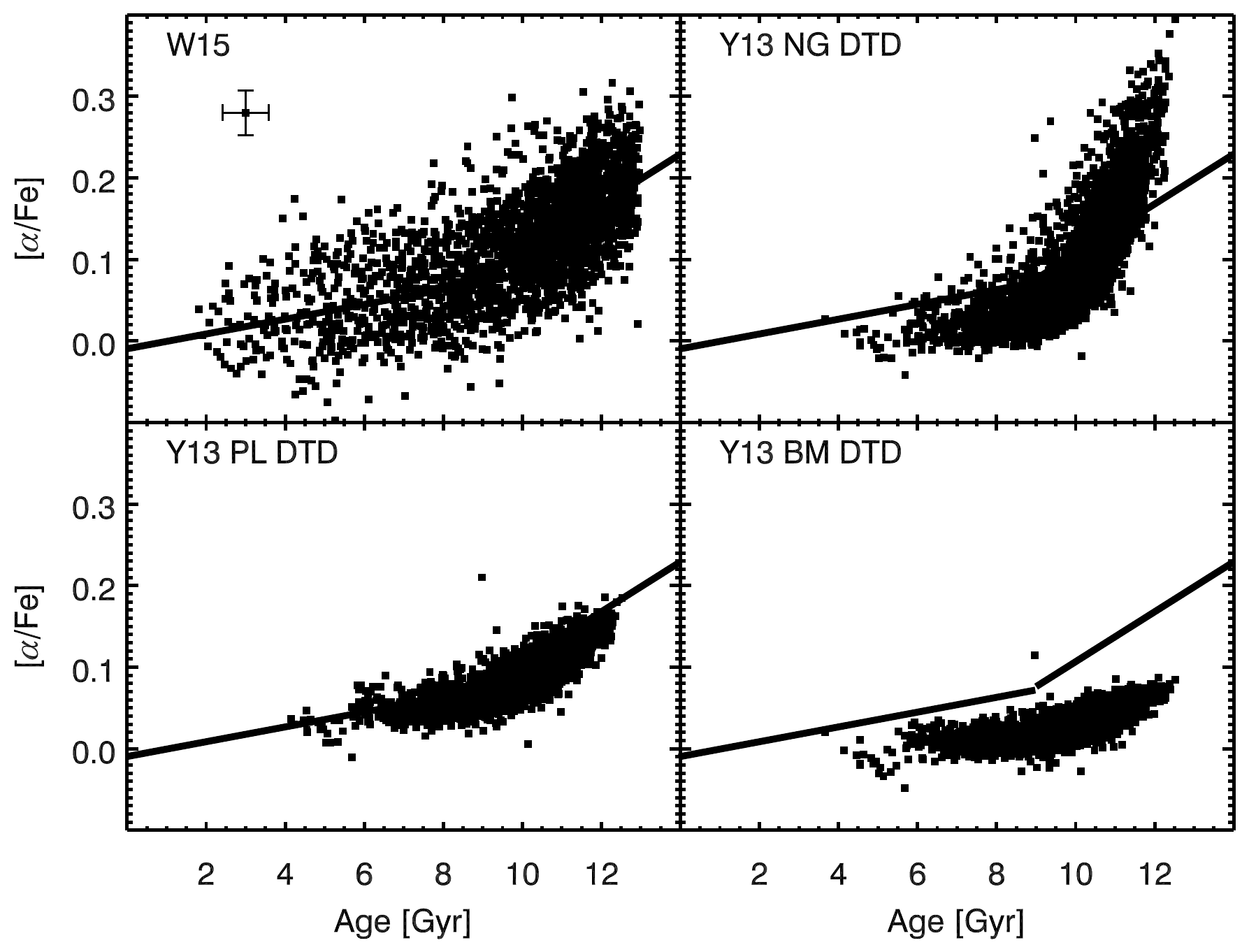}}
\end{center}
\caption[DTD]{Comparison of the correlations between age and \afe~for three different prescriptions for the delay time distribution 
of SNeIa in the Y13 model. \emph{Upper left panel:} The luminosity weighted average properties of early-type galaxies from 
W15. The solid line is a formal fit to the two regimes, separated at 9 Gyr and repeated in all the other panels. 
\emph{Upper right panel:} The luminosity weighted average properties of early-type galaxies in the semi-analytic model of 
Y13, using a Gaussian DTD. \emph{Lower left panel:} The luminosity weighted average properties of early-type galaxies 
in the semi-analytic model of Y13 using a power law DTD. \emph{Lower right panel:}  The luminosity weighted average 
properties of early-type galaxies in the semi-analytic model of Y13 using a bi-modal DTD. }
\label{f:obs_dtd}
\end{figure}

We repeat the relation between age and \afe~in Figure \ref{f:obs_dtd}, this time comparing it to the results from the Y13 model for different 
SNe Ia DTD. It seems fair to say that there is considerable debate in the literature on SNe Ia DTDs determined from direct observations of 
SNe and their host galaxies. Different authors claim different 
results with high certainty. In the hope of being representative we chose three DTDs, without any prejudice against other work. In all 
cases the delay time is denoted by $\tau$ and all DTDs are normalised to 1, such that 
\begin{equation}\label{eqn:DTDnormalisation}
\int^{\tau_{\textnormal{max}}}_{\tau_{\textnormal{min}}}\textnormal{DTD}(\tau)\ \textnormal{d}\tau=1. 
\end{equation}

\citet{mannucci06} found strong evidence for two different kinds of SNe Ia progenitors and proposed a bi-modal DTD: 
\begin{align} \label{eqn:ManDTD}
\nonumber \textnormal{log(DTD}_{\textnormal{BM}})& = \\
\bigg{\{} & \begin{array}{ll}
1.4-50(\textnormal{log}(\tau/\textnormal{yr})-7.7)^{2} & \textnormal{if } \tau < \tau_{0} \\
-0.8-0.9(\textnormal{log}(\tau/\textnormal{yr})-8.7)^{2} & \textnormal{if } \tau > \tau_{0} , 
\end{array}
\end{align}
where $\tau_0$ = 0.0851 Gyr separates the times where one or the other progenitor dominates the SN Ia rate. 

\citet{strolger04} on the other hand reject the double progenitor scenario "at the 99\% confidence level" and are able to describe 
their data by a narrow Gaussian DTD: 
\begin{equation} \label{eqn:ngDTD}
\textnormal{DTD}_{\textnormal{NG}} =
\frac{1}{\sqrt{2\pi\sigma_{\tau}^{2}}}\:e^{-\left( \tau-\tau_{c}\right)
^{2}/2\sigma_{\tau}^{2}}. 
\end{equation}
Here $\tau_c$ = 1 Gyr is the average delay time and $\sigma_{\tau} = 0.2 \tau_c$ Gyr is the width of the distribution. 

Finally, \citet{maoz12a} argue that the most recent data favour a power law DTD, which is described by 
\begin{equation} \label{eqn:plDTD}
\textnormal{DTD}_{\textnormal{PL}} = a(\tau/\textnormal{Gyr})^{-1.12}
\end{equation}
with normalization constant a = 0.15242 Gyr$^{-1}$. 

Figure \ref{f:obs_dtd} shows that the old-part slope of the age-\afe~relation is sensitive to the SNIa DTD. The power-law DTD is
clearly the best approximation of the data, while the two other DTDs fail at old ages. This result had been anticipated by earlier 
work. \citet{matteucci01} already show that a significant fraction of SNe Ia need to explode significantly before the 1 Gyr timescale 
often quoted for SNe Ia. Indeed for an instantaneous burst as assumed in the DTD they quote a typical timescale of very roughly 
50 Myr just as we are finding here. The fraction of SNe Ia to explode within 100 Myr after the burst of star formation has been further 
constrained by \citet{matteucci09} to be between 13\% and less than 30\%. It was estimated by Y13 to be $\sim$23\% for the 
power law DTD used here as well.  

We emphasize that the DTDs have been chosen directly from the literature on look back studies of SNe Ia explosion rates. These 
literature DTDs are naturally distinct and we have on purpose made no attempt to vary their functional parameters. 
For example, we could probably tweak the parameters of the 
bi-modal distribution to yield similar results to the power law DTD within our systematic measurement uncertainties. This would 
imply that the two DTDs are essentially the same as well, however. Note also that the downward re-normalization of the Y13 
data effected in Section \ref{s:obs_age_afe} is applied here as well, but does not in any way affect our conclusions. It is the 
shape of the age-\afe~correlation that allows us to diagnose the DTD, not the normalization of the \afe~values. 

\section{Discussion}
\label{s:discus} 

\subsection{Self-similarity or the independence on star formation history}
\label{s:selfsim}

The quantitative similarity of the relation age-\afe~presented in Figure \ref{f:obs_alp} for Milky Way stars and ETGs is not only not evident, 
it is even decidedly surprising. Indeed, the W15 and Y13 objects correspond to luminosity weighted average properties of 
massive galaxies, i.e. ensembles of more than $10^{10}$ stars, in galaxies that stopped forming the majority of their stars a 
long time ago. On the other hand the B14 and M13 objects are single stars in the solar neighborhood, i.e.~the thin disk of a 
nearly bulgeless disk galaxy that is still forming a few solar masses of stars every year (a much higher specific star formation 
rate than seen in present-day early-type galaxies). The apparent conundrum could be interpreted as follows: the B14 
age-\afe~relation for Milky Way stars could be tracing, down to z=0, a generic lookback time vs.~\afe~relation. If so, ETGs 
are simply galaxies that stopped forming stars somewhere earlier on that curve. An ETG sample with a range of ages 
would, therefore, populate an age-\afe~relation very similar to that of MW stars. 
Thus the most naive interpretation of the data would be that there is a common underlying age-\afe~relation that does not depend 
sensitively on the SFH of the galactic system. 

Such an interpretation is clearly oversimplified though. The thin disk curves in Figure 7 of \citet{minchev13} 
\citep[see also Figure 1 of][]{chiappini15} show that within the Milky Way the relation 
between age and \afe~is expected to depend on the radius of formation of the stars. The magenta line in that figure is very similar 
to our result here, but is the average result of chemo-dynamical evolution. In other words, the stellar line by itself mixes stars from 
different radii. The newly appeared pre-print by \citet[][their Figure 13]{anders16} would also seem to show significant scatter in the 
location of the knee of the \feh~vs.~\afe~plot, potentially tied to a variation in the age vs.~\afe~relation.  
We point out that, additionally to expected intrinsic variations of the age - \afe~relation in stellar systems, even for the Milky Ways stars, 
not all stars on this plot belong together in a causal sense. Indeed, 
the thick disk stars and the thin disk stars are distinct in their formation history, although formation times may overlap. This mirrors 
the statement made in W15 that the intermediate age stellar populations in ETGs are not causally connected to the old stellar populations. 

The chemical evolution models additionally allow to explore the plausibility of the two slope parametrization we have presented here for the 
observational data. Indeed, 
the two-infall model from \citet{chiappini97} essentially produces two different kinds of chemical evolutionary systems: 1) The thick disk, 
with high star formation efficiency and a short accretion timescale. 2) The thin disk, with lower star formation efficiency and an overall 
longer accretion timescale. The thin disk additionally has a varying accretion timescale with radius, being longer for larger radii in the 
Milky Way. Figure 2 of \citet{minchev13} shows the dependence of star formation history on radius within the Milky Way. Clearly, the 
center of the Milky Way experiences a star formation history that peaks at very early times, entirely opposite to the outermost radii, 
which have a very gentle increase of star formation rate over cosmic time. Yet, Figure 7 of  \citet{minchev13} shows that for the 
first three Gyr, the age-\afe~relation has a steep slope that varies only very slightly with radius, hence star formation history. 
The main effect of the varying star formation histories is in the slope of the age-\afe~relation at look back times less than 9 Gyr. 
In order to translate these insights to the ETGs, we additionally need to take into account that the number of stars on each of these 
tracks will vary widely: the longer the timescale of star formation (which corresponds to a larger radius in the Milky Way), the larger 
will be the fraction of stars on the tracks with an age less than 9 Gyr. The contrary is also true, i.e. a system with a very intense 
star formation burst at very early cosmic times will produce very few stars with very low \afe~values and ages less than 9 Gyr. 
Thus the net effect of averaging stellar populations from "chemical evolutionary systems" with varying star formation histories may 
be to drive towards a relation that is similar to the one shown here, or indeed to the average relation in the vicinity of the sun, as 
exemplified by the magenta line in Figure 7 of \citet{minchev13}. An example for how this averaging effect works at different radii 
in the disk of the Milky Way is shown in Figure 5 of \citet{minchev14}.

Thus, the apparent universality of the age-\afe~relation and its independence on the specific stellar assembly history for those 
two kinds of systems that we could test in the present contribution could be more than just a coincidence. It would rather be the 
expected average for complex stellar systems.  
It will be worthwhile and interesting to further study observationally whether this common relation exists for more stellar systems and to 
identify where and how different systems finally diverge, given small enough errorbars and as expected from chemical evolution models. 
Indeed, \citet{lehnert14} argue that "the low scatter in the \afe~as a function of age and the rapid decrease in \afe~with time suggests 
that mixing of metals was very efficient". It seems in light of the ETG data presented here that efficient mixing is not necessarily 
needed, if the age-\afe~relation is universal enough to apply to any star forming system with a mixture of stellar populations. 
It follows, however, that a lower mass limit to the validity of this relation must be expected, below which mixing arguments 
would have to be invoked to keep the relation universal. 

The apparent universality of the age-\afe~relation is not mirrored by the age-\feh~relations, which shows strong differences between 
the Milky Way stars and massive ETGs. 
Indeed, the W15 data show universally high \feh~values, while the B14 data show \feh~values that are lower by 1 dex for old stars. 
While the M13 model successfully reproduces the B14 data in this figure, the Y13 model shows an offset in \feh~as compared to W15. 
In-depth discussion of this offset is beyond the scope of the present contribution. A tentative solution to be explored elsewhere is 
that the negative offset and flatter slope of the relation is a mass effect. The typical $\sim$6 Gyr old ETG in the Y13 model is less 
massive than that in the W15 sample, even though the Y13 sub-sample used here is mass selected. This sample selection effect 
could be partially caused by the selection on signal-to-noise for the observational data points, as discussed in W15. 
However, the overall trend is the same as shown in the W15 data, i.e.~an anti-correlation between age and \feh. 

Note that beyond the mean relations, the scatter of the age-\afe~and age-\feh~relations may contain physical insight if the error bars can be 
driven further down. Discussing the case of the ETGs, the oldest ellipticals with ages $>$ 11 Gyr and log(M*) $>$ 11.5 \msun at z = 0, 
did not have time to enrich heavily in Iron, therefore they should show very small scatter in \afe. Some slightly younger massive 
ellipticals would have had slightly longer star formation timescales, hence lowering the \afe~and increasing their \feh. Other slightly younger 
massive ellipticals would have started forming their stars later in the history of the universe, leading them to show overall higher 
\afe~and lower \feh~at the same age. The scatter in \afe~and \feh~may thus turn out to be a good diagnostic of the time of onset of 
star formation, a quantity that has eluded observational constraints from galactic archeology for any galaxy we cannot resolve in 
single stars. 

\subsection{Constraints on the Delay Time Distribution of SNeIa}
\label{s:dtd}

In the last Section \ref{s:selfsim} we have stated that the shape of the age-\afe~relation is relatively independent on the specific star 
formation history within the two stellar systems probed here. As shown in Figure \ref{f:obs_dtd} the power-law SNe Ia DTD reproduces
the W15 results best. A very similar DTD was inferred from earlier constraints on chemical evolution models using Milky Way 
data in \citet{matteucci01}, with a similar peak in SNe Ia rate at 40-50 Myr after the burst. This DTD was also 
preferred in Y13 for the M*-[O/Fe] relation, and for the oxygen enhancement in MW disc stars. While the use of the power-law DTD
is consistent with the literature on direct supernova observations \citep{maoz12b}, as shown by \citet{bonaparte13} it turns out that the 
tighter constraint on the DTD comes from chemical evolution arguments as used in the present paper for ETGs and earlier in 
\citet{matteucci06} and \citet{matteucci09} for the Milky Way. 

When looking at the different DTDs it turns out that the slope of the relation below 9 Gyrs is roughly the same for all DTDs. While the 
normalization changes slightly, for reasons discussed above we do not consider the normalization a robust discriminant. However, 
the steepness of the slope for the $>9$ Gyr population is sensitive to the DTD. We emphasize that this is also the part of the age-\afe~ 
relation that tends to look more universal. A DTD that produces \emph{fewer} prompt SNe-Ia 
exhibit a \emph{steeper} slope, because of the higher starting \afe~values. This can be understood through the luminosity weighted 
average nature of the plotted quantities. If there are more prompt SNeIa, the \afe~of the old stars will still be a mix of high and low 
\afe~stars. Thus, even for the oldest galaxies the \afe~will be low. If there are less prompt SNe Ia, the oldest galaxies will be 
dominated by high \afe~stars. On the other hand, for all DTDs, 3 Gyr after the onset of star formation (i.e.~around 9 to 10 Gyr 
lookback time), the \afe~ratio will have reached the same, low value around 0.05. 

This point is complimentary to the dependence of the slope of the M*-[O/Fe] relation on the DTD that was already discussed in
Y13. Older model galaxies have shorter star formation timescales, and the [O/Fe] at z=0 of the oldest galaxies will be higher 
(i.e.~closer to the ratio produced by low-metallicity SNe II) for DTDs with smaller prompt components.

Finally, we note as a caveat that we have neglected IMF variations for the arguments presented here. As very recently pointed 
out again in \citet{martin-navarro15}, a change of IMF does have an effect on \afe~evolution and therefore could potentially 
affect the inferences concerning the DTD. On the other hand, changes in the IMF result mostly in changes of the \afe~plateau 
value and not in the actual evolution of \afe~with age \citep{romano05}. Also, O is more affected than Mg, because 
O yields change more significantly with stellar mass than those of Mg.


\subsection{Other combinations of DTD and SFH that fit the data}
\label{s:other comb}

\citet{snaith14} showed that the stellar data of H13 can be fitted with a significantly different DTD. Their DTD is based on a physical 
model with a single degenerate progenitor \citep{kawata03}. It is bi-modal, with one component due to main sequence mass 
donors and a more delayed component due to red giant mass donors. However, none of those is a 'prompt' component in the 
classical sense, as no SN Ia explodes before 0.7 Gyr after star formation. In both the bi-modal and the power law DTD used 
in the present work, the first SNe Ia explode after 0.035 Gyr and about half of all SNe Ia explode before 0.4 Gyr. Our chosen 
minimum delay time reflects the lifetime of an 8 \msun star, the most massive secondary companion normally assumed in 
SNIa progenitor models \citet{matteucci86}, \citet{greggio05}, \citet{matteucci06}, and \citet{matteucci09}. 
It also allows us to meet observational constraints on the SNIa rate \citep{brandt10, maoz10}. 

The difference in DTD choice between \citet{snaith14} and this work \citep[and e.g.~][]{matteucci01} has a consequence on the old-part 
slope of the age-\afe~relation. In their case, the starting value of \afe~($\sim$13 Gyr ago) is always the same, but the end value of \afe~(at 
$\sim$9 Gyr) depends on the star formation history. In our case, the end value is always the same, but the starting value 
depends on the DTD (i.e.~number of prompt SNeIa with delay times between 35 and 100 Myr). Therefore, we can both 
obtain very similar old-part slopes, but for very different reasons.
Distinguishing between the \citet{snaith14} results and ours thus hinges on the SN Ia rates at delay times between 35 and 100 Myr. 

\section{Conclusions}
\label{s:concl} 

We have compared the age-\afe~relation between ETGs and the solar neighborhood, for data and models. We find that the relation 
is quantitatively the same, and that both Milky Way and early-type galaxy data require a DTD with a small prompt component 
(<30\% of SNe-Ia exploding within 100 Myr). For example, a power-law DTD, such as those commonly derived from observations of 
the SN-Ia rate, matches this requirement. We also suggest that the observed scatter in the age-\afe~relation for ETGs could be 
driven by differences in the onset of star formation in those systems. For the actually existing range of galactic systems and 
therefore star formation histories studied in the present paper, the age-\afe~relation is self-similar on widely different scales.  
A tentative explanation for this seeming universality 
of the age-\afe~relation is that is results from averaging of different stellar populations with varying chemical evolution histories. 
It thus does not seem to be a useful tool to understand the 
star formation histories of galaxies, contrary to the more widely used \feh-\afe~relations. 

\begin{acknowledgements}

We acknowledge the report from an anonymous referee, that helped to significantly sharpen the arguments presented in this paper. 
CJW and PC acknowledge support through the Marie Curie Career Integration Grant 303912. 
RMY acknowledges support through the Sofia Kovalevskaja Award to P. Schady from the Alexander von Humboldt Foundation of Germany.
SC acknowledges support from the European Research Council via an Advanced Grant under grant agreement no. 321323-NEOGAL. 
GB acknowledges support for this work from the National Autonomous University of M\'exico (UNAM), through grant PAPIIT IG100115.

Funding for the SDSS and SDSS-II has been provided by the Alfred P. Sloan Foundation, the Participating Institutions, the National Science Foundation, the U.S. Department of Energy, the National Aeronautics and Space Administration, the Japanese Monbukagakusho, the Max Planck Society, and the Higher Education Funding Council for England. The SDSS Web Site is http://www.sdss.org/.

The SDSS is managed by the Astrophysical Research Consortium for the Participating Institutions. The Participating Institutions are the American Museum of Natural History, Astrophysical Institute Potsdam, University of Basel, University of Cambridge, Case Western Reserve University, University of Chicago, Drexel University, Fermilab, the Institute for Advanced Study, the Japan Participation Group, Johns Hopkins University, the Joint Institute for Nuclear Astrophysics, the Kavli Institute for Particle Astrophysics and Cosmology, the Korean Scientist Group, the Chinese Academy of Sciences (LAMOST), Los Alamos National Laboratory, the Max-Planck-Institute for Astronomy (MPIA), the Max-Planck-Institute for Astrophysics (MPA), New Mexico State University, Ohio State University, University of Pittsburgh, University of Portsmouth, Princeton University, the United States Naval Observatory, and the University of Washington.

\end{acknowledgements}

\bibliographystyle{aa}
\bibliography{selfsimilar}

\begin{thebibliography}{60}
\expandafter\ifx\csname natexlab\endcsname\relax\def\natexlab#1{#1}\fi

\bibitem[{{Abazajian} {et~al.}(2009){Abazajian}, {Adelman-McCarthy},
  {Ag{\"u}eros}, {Allam}, {Allende Prieto}, {An}, {Anderson}, {Anderson},
  {Annis}, {Bahcall}, {Bailer-Jones}, {Barentine}, {Bassett}, {Becker},
  {Beers}, {Bell}, {Belokurov}, {Berlind}, {Berman}, {Bernardi}, {Bickerton},
  {Bizyaev}, {Blakeslee}, {Blanton}, {Bochanski}, {Boroski}, {Brewington},
  {Brinchmann}, {Brinkmann}, {Brunner}, {Budav{\'a}ri}, {Carey}, {Carliles},
  {Carr}, {Castander}, {Cinabro}, {Connolly}, {Csabai}, {Cunha}, {Czarapata},
  {Davenport}, {de Haas}, {Dilday}, {Doi}, {Eisenstein}, {Evans}, {Evans},
  {Fan}, {Friedman}, {Frieman}, {Fukugita}, {G{\"a}nsicke}, {Gates},
  {Gillespie}, {Gilmore}, {Gonzalez}, {Gonzalez}, {Grebel}, {Gunn},
  {Gy{\"o}ry}, {Hall}, {Harding}, {Harris}, {Harvanek}, {Hawley}, {Hayes},
  {Heckman}, {Hendry}, {Hennessy}, {Hindsley}, {Hoblitt}, {Hogan}, {Hogg},
  {Holtzman}, {Hyde}, {Ichikawa}, {Ichikawa}, {Im}, {Ivezi{\'c}}, {Jester},
  {Jiang}, {Johnson}, {Jorgensen}, {Juri{\'c}}, {Kent}, {Kessler}, {Kleinman},
  {Knapp}, {Konishi}, {Kron}, {Krzesinski}, {Kuropatkin}, {Lampeitl},
  {Lebedeva}, {Lee}, {Lee}, {Leger}, {L{\'e}pine}, {Li}, {Lima}, {Lin}, {Long},
  {Loomis}, {Loveday}, {Lupton}, {Magnier}, {Malanushenko}, {Malanushenko},
  {Mandelbaum}, {Margon}, {Marriner}, {Mart{\'{\i}}nez-Delgado}, {Matsubara},
  {McGehee}, {McKay}, {Meiksin}, {Morrison}, {Mullally}, {Munn}, {Murphy},
  {Nash}, {Nebot}, {Neilsen}, {Newberg}, {Newman}, {Nichol}, {Nicinski},
  {Nieto-Santisteban}, {Nitta}, {Okamura}, {Oravetz}, {Ostriker}, {Owen},
  {Padmanabhan}, {Pan}, {Park}, {Pauls}, {Peoples}, {Percival}, {Pier}, {Pope},
  {Pourbaix}, {Price}, {Purger}, {Quinn}, {Raddick}, {Fiorentin}, {Richards},
  {Richmond}, {Riess}, {Rix}, {Rockosi}, {Sako}, {Schlegel}, {Schneider},
  {Scholz}, {Schreiber}, {Schwope}, {Seljak}, {Sesar}, {Sheldon}, {Shimasaku},
  {Sibley}, {Simmons}, {Sivarani}, {Smith}, {Smith}, {Smol{\v c}i{\'c}},
  {Snedden}, {Stebbins}, {Steinmetz}, {Stoughton}, {Strauss}, {Subba Rao},
  {Suto}, {Szalay}, {Szapudi}, {Szkody}, {Tanaka}, {Tegmark}, {Teodoro},
  {Thakar}, {Tremonti}, {Tucker}, {Uomoto}, {Vanden Berk}, {Vandenberg},
  {Vidrih}, {Vogeley}, {Voges}, {Vogt}, {Wadadekar}, {Watters}, {Weinberg},
  {West}, {White}, {Wilhite}, {Wonders}, {Yanny}, {Yocum}, {York}, {Zehavi},
  {Zibetti}, \& {Zucker}}]{abazajian09}
{Abazajian}, K.~N., {Adelman-McCarthy}, J.~K., {Ag{\"u}eros}, M.~A., {et~al.}
  2009, \apjs, 182, 543

\bibitem[{{Adibekyan} {et~al.}(2012){Adibekyan}, {Sousa}, {Santos}, {Delgado
  Mena}, {Gonz{\'a}lez Hern{\'a}ndez}, {Israelian}, {Mayor}, \&
  {Khachatryan}}]{adibekyan12}
{Adibekyan}, V.~Z., {Sousa}, S.~G., {Santos}, N.~C., {et~al.} 2012, \aap, 545,
  A32

\bibitem[{{Anders} \& {Grevesse}(1989)}]{anders89}
{Anders}, E. \& {Grevesse}, N. 1989, \gca, 53, 197

\bibitem[{{Anders} {et~al.}(2016){Anders}, {Chiappini}, {Rodrigues}, {Miglio},
  {Montalb{\'a}n}, {Mosser}, {Girardi}, {Valentini}, {Noels}, {Morel},
  {Johnson}, {Schultheis}, {Baudin}, {de Assis Peralta}, {Hekker},
  {Theme{\ss}l}, {Kallinger}, {Garc{\'{\i}}a}, {Mathur}, {Baglin}, {Santiago},
  {Martig}, {Minchev}, {Steinmetz}, {da Costa}, {Maia}, {Allende Prieto},
  {Cunha}, {Beers}, {Epstein}, {Garc{\'{\i}}a P{\'e}rez},
  {Garc{\'{\i}}a-Hern{\'a}ndez}, {Harding}, {Holtzman}, {Majewski},
  {M{\'e}sz{\'a}ros}, {Nidever}, {Pan}, {Pinsonneault}, {Schiavon},
  {Schneider}, {Shetrone}, {Stassun}, {Zamora}, \& {Zasowski}}]{anders16}
{Anders}, F., {Chiappini}, C., {Rodrigues}, T.~S., {et~al.} 2016, ArXiv
  e-prints

\bibitem[{{Bensby} {et~al.}(2014){Bensby}, {Feltzing}, \& {Oey}}]{bensby14}
{Bensby}, T., {Feltzing}, S., \& {Oey}, M.~S. 2014, \aap, 562, A71

\bibitem[{{Bergemann} {et~al.}(2014){Bergemann}, {Ruchti}, {Serenelli},
  {Feltzing}, {Alves-Brito}, {Asplund}, {Bensby}, {Gruyters}, {Heiter},
  {Hourihane}, {Korn}, {Lind}, {Marino}, {Jofre}, {Nordlander}, {Ryde},
  {Worley}, {Gilmore}, {Randich}, {Ferguson}, {Jeffries}, {Micela},
  {Negueruela}, {Prusti}, {Rix}, {Vallenari}, {Alfaro}, {Allende Prieto},
  {Bragaglia}, {Koposov}, {Lanzafame}, {Pancino}, {Recio-Blanco}, {Smiljanic},
  {Walton}, {Costado}, {Franciosini}, {Hill}, {Lardo}, {de Laverny}, {Magrini},
  {Maiorca}, {Masseron}, {Morbidelli}, {Sacco}, {Kordopatis}, \& {Tautvai{\v
  s}ien{\.e}}}]{bergemann14}
{Bergemann}, M., {Ruchti}, G.~R., {Serenelli}, A., {et~al.} 2014, \aap, 565,
  A89

\bibitem[{{Bonaparte} {et~al.}(2013){Bonaparte}, {Matteucci}, {Recchi},
  {Spitoni}, {Pipino}, \& {Grieco}}]{bonaparte13}
{Bonaparte}, I., {Matteucci}, F., {Recchi}, S., {et~al.} 2013, \mnras, 435,
  2460

\bibitem[{{Boylan-Kolchin} {et~al.}(2009){Boylan-Kolchin}, {Springel}, {White},
  {Jenkins}, \& {Lemson}}]{boylan-kolchin09}
{Boylan-Kolchin}, M., {Springel}, V., {White}, S.~D.~M., {Jenkins}, A., \&
  {Lemson}, G. 2009, \mnras, 398, 1150

\bibitem[{{Brandt} {et~al.}(2010){Brandt}, {Tojeiro}, {Aubourg}, {Heavens},
  {Jimenez}, \& {Strauss}}]{brandt10}
{Brandt}, T.~D., {Tojeiro}, R., {Aubourg}, {\'E}., {et~al.} 2010, \aj, 140, 804

\bibitem[{{Casagrande} {et~al.}(2011){Casagrande}, {Sch{\"o}nrich}, {Asplund},
  {Cassisi}, {Ram{\'{\i}}rez}, {Mel{\'e}ndez}, {Bensby}, \&
  {Feltzing}}]{casagrande11}
{Casagrande}, L., {Sch{\"o}nrich}, R., {Asplund}, M., {et~al.} 2011, \aap, 530,
  A138

\bibitem[{{Chiappini} {et~al.}(2015){Chiappini}, {Anders}, {Rodrigues},
  {Miglio}, {Montalb{\'a}n}, {Mosser}, {Girardi}, {Valentini}, {Noels},
  {Morel}, {Minchev}, {Steinmetz}, {Santiago}, {Schultheis}, {Martig}, {da
  Costa}, {Maia}, {Allende Prieto}, {de Assis Peralta}, {Hekker},
  {Theme{\ss}l}, {Kallinger}, {Garc{\'{\i}}a}, {Mathur}, {Baudin}, {Beers},
  {Cunha}, {Harding}, {Holtzman}, {Majewski}, {M{\'e}sz{\'a}ros}, {Nidever},
  {Pan}, {Schiavon}, {Shetrone}, {Schneider}, \& {Stassun}}]{chiappini15}
{Chiappini}, C., {Anders}, F., {Rodrigues}, T.~S., {et~al.} 2015, \aap, 576,
  L12

\bibitem[{{Chiappini} {et~al.}(1997){Chiappini}, {Matteucci}, \&
  {Gratton}}]{chiappini97}
{Chiappini}, C., {Matteucci}, F., \& {Gratton}, R. 1997, \apj, 477, 765

\bibitem[{{Coelho} {et~al.}(2007){Coelho}, {Bruzual}, {Charlot}, {Weiss},
  {Barbuy}, \& {Ferguson}}]{coelho07}
{Coelho}, P., {Bruzual}, G., {Charlot}, S., {et~al.} 2007, \mnras, 382, 498

\bibitem[{{Edvardsson} {et~al.}(1993){Edvardsson}, {Andersen}, {Gustafsson},
  {Lambert}, {Nissen}, \& {Tomkin}}]{edvardsson93}
{Edvardsson}, B., {Andersen}, J., {Gustafsson}, B., {et~al.} 1993, \aap, 275,
  101

\bibitem[{{Feltzing} {et~al.}(2001){Feltzing}, {Holmberg}, \&
  {Hurley}}]{feltzing01}
{Feltzing}, S., {Holmberg}, J., \& {Hurley}, J.~R. 2001, \aap, 377, 911

\bibitem[{{Gallazzi} {et~al.}(2006){Gallazzi}, {Charlot}, {Brinchmann}, \&
  {White}}]{gallazzi06}
{Gallazzi}, A., {Charlot}, S., {Brinchmann}, J., \& {White}, S.~D.~M. 2006,
  \mnras, 370, 1106

\bibitem[{{Gilmore} {et~al.}(2012){Gilmore}, {Randich}, {Asplund}, {Binney},
  {Bonifacio}, {Drew}, {Feltzing}, {Ferguson}, {Jeffries}, {Micela}, \&
  et~al.}]{gilmore12}
{Gilmore}, G., {Randich}, S., {Asplund}, M., {et~al.} 2012, The Messenger, 147,
  25

\bibitem[{{Graves} {et~al.}(2010){Graves}, {Faber}, \& {Schiavon}}]{graves10}
{Graves}, G.~J., {Faber}, S.~M., \& {Schiavon}, R.~P. 2010, \apj, 721, 278

\bibitem[{{Greggio}(2005)}]{greggio05}
{Greggio}, L. 2005, \aap, 441, 1055

\bibitem[{{Grevesse} \& {Sauval}(1998)}]{grevesse98}
{Grevesse}, N. \& {Sauval}, A.~J. 1998, \ssr, 85, 161

\bibitem[{{Guo} {et~al.}(2011){Guo}, {White}, {Boylan-Kolchin}, {De Lucia},
  {Kauffmann}, {Lemson}, {Li}, {Springel}, \& {Weinmann}}]{guo11}
{Guo}, Q., {White}, S., {Boylan-Kolchin}, M., {et~al.} 2011, \mnras, 413, 101

\bibitem[{{Haywood} {et~al.}(2013){Haywood}, {Di Matteo}, {Lehnert}, {Katz}, \&
  {G{\'o}mez}}]{haywood13}
{Haywood}, M., {Di Matteo}, P., {Lehnert}, M.~D., {Katz}, D., \& {G{\'o}mez},
  A. 2013, \aap, 560, A109

\bibitem[{{Heiter} {et~al.}(2015){Heiter}, {Jofr{\'e}}, {Gustafsson}, {Korn},
  {Soubiran}, \& {Th{\'e}venin}}]{heiter15}
{Heiter}, U., {Jofr{\'e}}, P., {Gustafsson}, B., {et~al.} 2015, \aap, 582, A49

\bibitem[{{Holmberg} {et~al.}(2007){Holmberg}, {Nordstr{\"o}m}, \&
  {Andersen}}]{holmberg07}
{Holmberg}, J., {Nordstr{\"o}m}, B., \& {Andersen}, J. 2007, \aap, 475, 519

\bibitem[{{Holmberg} {et~al.}(2009){Holmberg}, {Nordstr{\"o}m}, \&
  {Andersen}}]{holmberg09}
{Holmberg}, J., {Nordstr{\"o}m}, B., \& {Andersen}, J. 2009, \aap, 501, 941

\bibitem[{{Jofr{\'e}} {et~al.}(2014){Jofr{\'e}}, {Heiter}, {Soubiran},
  {Blanco-Cuaresma}, {Worley}, {Pancino}, {Cantat-Gaudin}, {Magrini},
  {Bergemann}, {Gonz{\'a}lez Hern{\'a}ndez}, {Hill}, {Lardo}, {de Laverny},
  {Lind}, {Masseron}, {Montes}, {Mucciarelli}, {Nordlander}, {Recio Blanco},
  {Sobeck}, {Sordo}, {Sousa}, {Tabernero}, {Vallenari}, \& {Van Eck}}]{jofre14}
{Jofr{\'e}}, P., {Heiter}, U., {Soubiran}, C., {et~al.} 2014, \aap, 564, A133

\bibitem[{{Johansson} {et~al.}(2012){Johansson}, {Thomas}, \&
  {Maraston}}]{johansson12}
{Johansson}, J., {Thomas}, D., \& {Maraston}, C. 2012, \mnras, 421, 1908

\bibitem[{{J{\o}rgensen}(1999)}]{jorgensen99}
{J{\o}rgensen}, I. 1999, \mnras, 306, 607

\bibitem[{{Kawata} \& {Gibson}(2003)}]{kawata03}
{Kawata}, D. \& {Gibson}, B.~K. 2003, \mnras, 346, 135

\bibitem[{{Kuntschner} {et~al.}(2010){Kuntschner}, {Emsellem}, {Bacon},
  {Cappellari}, {Davies}, {de Zeeuw}, {Falc{\'o}n-Barroso}, {Krajnovi{\'c}},
  {McDermid}, {Peletier}, {Sarzi}, {Shapiro}, {van den Bosch}, \& {van de
  Ven}}]{kuntschner10}
{Kuntschner}, H., {Emsellem}, E., {Bacon}, R., {et~al.} 2010, \mnras, 408, 97

\bibitem[{{Lehnert} {et~al.}(2014){Lehnert}, {Di Matteo}, {Haywood}, \&
  {Snaith}}]{lehnert14}
{Lehnert}, M.~D., {Di Matteo}, P., {Haywood}, M., \& {Snaith}, O.~N. 2014,
  \apjl, 789, L30

\bibitem[{{Mannucci} {et~al.}(2006){Mannucci}, {Della Valle}, \&
  {Panagia}}]{mannucci06}
{Mannucci}, F., {Della Valle}, M., \& {Panagia}, N. 2006, \mnras, 370, 773

\bibitem[{{Maoz} \& {Badenes}(2010)}]{maoz10}
{Maoz}, D. \& {Badenes}, C. 2010, \mnras, 407, 1314

\bibitem[{{Maoz} \& {Mannucci}(2012b)}]{maoz12b}
{Maoz}, D. \& {Mannucci}, F. 2012b, \pasa, 29, 447

\bibitem[{{Maoz} {et~al.}(2012a){Maoz}, {Mannucci}, \& {Brandt}}]{maoz12a}
{Maoz}, D., {Mannucci}, F., \& {Brandt}, T.~D. 2012a, \mnras, 426, 3282

\bibitem[{{Martig} {et~al.}(2012){Martig}, {Bournaud}, {Croton}, {Dekel}, \&
  {Teyssier}}]{martig12}
{Martig}, M., {Bournaud}, F., {Croton}, D.~J., {Dekel}, A., \& {Teyssier}, R.
  2012, \apj, 756, 26

\bibitem[{{Martig} {et~al.}(2015){Martig}, {Rix}, {Silva Aguirre}, {Hekker},
  {Mosser}, {Elsworth}, {Bovy}, {Stello}, {Anders}, {Garc{\'{\i}}a}, {Tayar},
  {Rodrigues}, {Basu}, {Carrera}, {Ceillier}, {Chaplin}, {Chiappini},
  {Frinchaboy}, {Garc{\'{\i}}a-Hern{\'a}ndez}, {Hearty}, {Holtzman}, {Johnson},
  {Majewski}, {Mathur}, {M{\'e}sz{\'a}ros}, {Miglio}, {Nidever}, {Pan},
  {Pinsonneault}, {Schiavon}, {Schneider}, {Serenelli}, {Shetrone}, \&
  {Zamora}}]{martig15}
{Martig}, M., {Rix}, H.-W., {Silva Aguirre}, V., {et~al.} 2015, \mnras, 451,
  2230

\bibitem[{{Mart{\'{\i}}n-Navarro}(2015)}]{martin-navarro15}
{Mart{\'{\i}}n-Navarro}, I. 2015, ArXiv e-prints

\bibitem[{{Matteucci} \& {Greggio}(1986)}]{matteucci86}
{Matteucci}, F. \& {Greggio}, L. 1986, \aap, 154, 279

\bibitem[{{Matteucci} {et~al.}(2006){Matteucci}, {Panagia}, {Pipino},
  {Mannucci}, {Recchi}, \& {Della Valle}}]{matteucci06}
{Matteucci}, F., {Panagia}, N., {Pipino}, A., {et~al.} 2006, \mnras, 372, 265

\bibitem[{{Matteucci} \& {Recchi}(2001)}]{matteucci01}
{Matteucci}, F. \& {Recchi}, S. 2001, \apj, 558, 351

\bibitem[{{Matteucci} {et~al.}(2009){Matteucci}, {Spitoni}, {Recchi}, \&
  {Valiante}}]{matteucci09}
{Matteucci}, F., {Spitoni}, E., {Recchi}, S., \& {Valiante}, R. 2009, \aap,
  501, 531

\bibitem[{{Minchev} {et~al.}(2013){Minchev}, {Chiappini}, \&
  {Martig}}]{minchev13}
{Minchev}, I., {Chiappini}, C., \& {Martig}, M. 2013, \aap, 558, A9

\bibitem[{{Minchev} {et~al.}(2014){Minchev}, {Chiappini}, \&
  {Martig}}]{minchev14}
{Minchev}, I., {Chiappini}, C., \& {Martig}, M. 2014, \aap, 572, A92

\bibitem[{{Nordstr{\"o}m} {et~al.}(2004){Nordstr{\"o}m}, {Mayor}, {Andersen},
  {Holmberg}, {Pont}, {J{\o}rgensen}, {Olsen}, {Udry}, \&
  {Mowlavi}}]{nordstrom04}
{Nordstr{\"o}m}, B., {Mayor}, M., {Andersen}, J., {et~al.} 2004, \aap, 418, 989

\bibitem[{{Portinari} {et~al.}(1998){Portinari}, {Chiosi}, \&
  {Bressan}}]{portinari98}
{Portinari}, L., {Chiosi}, C., \& {Bressan}, A. 1998, \aap, 334, 505

\bibitem[{{Randich} {et~al.}(2013){Randich}, {Gilmore}, \& {Gaia-ESO
  Consortium}}]{randich13}
{Randich}, S., {Gilmore}, G., \& {Gaia-ESO Consortium}. 2013, The Messenger,
  154, 47

\bibitem[{{Romano} {et~al.}(2005){Romano}, {Chiappini}, {Matteucci}, \&
  {Tosi}}]{romano05}
{Romano}, D., {Chiappini}, C., {Matteucci}, F., \& {Tosi}, M. 2005, \aap, 430,
  491

\bibitem[{{Serenelli} {et~al.}(2013){Serenelli}, {Bergemann}, {Ruchti}, \&
  {Casagrande}}]{serenelli13}
{Serenelli}, A.~M., {Bergemann}, M., {Ruchti}, G., \& {Casagrande}, L. 2013,
  \mnras, 429, 3645

\bibitem[{{Snaith} {et~al.}(2014){Snaith}, {Haywood}, {Di Matteo}, {Lehnert},
  {Combes}, {Katz}, \& {G{\'o}mez}}]{snaith14}
{Snaith}, O.~N., {Haywood}, M., {Di Matteo}, P., {et~al.} 2014, \apjl, 781, L31

\bibitem[{{Springel} {et~al.}(2005){Springel}, {White}, {Jenkins}, {Frenk},
  {Yoshida}, {Gao}, {Navarro}, {Thacker}, {Croton}, {Helly}, {Peacock}, {Cole},
  {Thomas}, {Couchman}, {Evrard}, {Colberg}, \& {Pearce}}]{springel05}
{Springel}, V., {White}, S.~D.~M., {Jenkins}, A., {et~al.} 2005, \nat, 435, 629

\bibitem[{{Springel} {et~al.}(2001){Springel}, {White}, {Tormen}, \&
  {Kauffmann}}]{springel01}
{Springel}, V., {White}, S.~D.~M., {Tormen}, G., \& {Kauffmann}, G. 2001,
  \mnras, 328, 726

\bibitem[{{Strolger} {et~al.}(2004){Strolger}, {Riess}, {Dahlen}, {Livio},
  {Panagia}, {Challis}, {Tonry}, {Filippenko}, {Chornock}, {Ferguson},
  {Koekemoer}, {Mobasher}, {Dickinson}, {Giavalisco}, {Casertano}, {Hook},
  {Blondin}, {Leibundgut}, {Nonino}, {Rosati}, {Spinrad}, {Steidel}, {Stern},
  {Garnavich}, {Matheson}, {Grogin}, {Hornschemeier}, {Kretchmer}, {Laidler},
  {Lee}, {Lucas}, {de Mello}, {Moustakas}, {Ravindranath}, {Richardson}, \&
  {Taylor}}]{strolger04}
{Strolger}, L.-G., {Riess}, A.~G., {Dahlen}, T., {et~al.} 2004, \apj, 613, 200

\bibitem[{{Thomas} {et~al.}(2005){Thomas}, {Maraston}, {Bender}, \& {Mendes de
  Oliveira}}]{thomas05}
{Thomas}, D., {Maraston}, C., {Bender}, R., \& {Mendes de Oliveira}, C. 2005,
  \apj, 621, 673

\bibitem[{{Thomas} {et~al.}(2010){Thomas}, {Maraston}, {Schawinski}, {Sarzi},
  \& {Silk}}]{thomas10}
{Thomas}, D., {Maraston}, C., {Schawinski}, K., {Sarzi}, M., \& {Silk}, J.
  2010, \mnras, 404, 1775

\bibitem[{{Tinsley}(1979)}]{tinsley79}
{Tinsley}, B.~M. 1979, \apj, 229, 1046

\bibitem[{{Twarog}(1980)}]{twarog80}
{Twarog}, B.~A. 1980, \apj, 242, 242

\bibitem[{{Walcher} {et~al.}(2009){Walcher}, {Coelho}, {Gallazzi}, \&
  {Charlot}}]{walcher09}
{Walcher}, C.~J., {Coelho}, P., {Gallazzi}, A., \& {Charlot}, S. 2009, \mnras,
  L275

\bibitem[{{Walcher} {et~al.}(2015){Walcher}, {Coelho}, {Gallazzi}, {Bruzual},
  {Charlot}, \& {Chiappini}}]{walcher15}
{Walcher}, C.~J., {Coelho}, P.~R.~T., {Gallazzi}, A., {et~al.} 2015, \aap, 582,
  A46

\bibitem[{{Yates} {et~al.}(2013){Yates}, {Henriques}, {Thomas}, {Kauffmann},
  {Johansson}, \& {White}}]{yates13}
{Yates}, R.~M., {Henriques}, B., {Thomas}, P.~A., {et~al.} 2013, \mnras, 435,
  3500

\end{thebibliography}

\end{document}